\documentclass[aps,prd,preprint,nofootinbib]{revtex4}
\usepackage{amsfonts}
\usepackage{mathrsfs}
\usepackage{graphicx}
\usepackage{amsmath}
\usepackage{amssymb}
\usepackage{epsfig}
\usepackage{graphicx}
\usepackage{slashed}
\usepackage{color}
\usepackage{multirow}
\usepackage{tabularx} 

\usepackage{ulem}
\usepackage{adjustbox}

\usepackage{stmaryrd}
\newcommand{\bqa}{\begin{eqnarray}}
	\newcommand{\eqa}{\end{eqnarray}}
\newcommand{\beq}{\begin{equation}}
	\newcommand{\eeq}{\end{equation}}
\allowdisplaybreaks[1]

\graphicspath{{fig/}{dia/}} \DeclareGraphicsExtensions{.eps}
\begin{document}
	
	\title{Heavy-Flavor-Conserving Hadronic Weak Decays of Charmed and Bottom Baryons: an Update
	}

	\author {Hai-Yang Cheng$^1$, Chia-Wei Liu$^2$ and Fanrong Xu$^3$ }
	\affiliation{	
		$^1$Institute of Physics, Academia Sinica,	Taipei, Taiwan 115, Republic of China\\
		$^2$School of Fundamental Physics and Mathematical Sciences, Hangzhou Institute for Advanced Study, UCAS, Hangzhou 310024, China\\
		$^3$ Department of Physics, Jinan University\\ Guangzhou 510632, People’s Republic of China
	}
	\date{\today}

	\begin{abstract}
		This paper serves as an update of the previous work entitled ``Heavy-Flavor-Conserving Hadronic Weak Decays of Charmed and Bottom Baryons". We make an improvement on the bag wave functions by removing the center-of-mass motion of the bag. 
		All the baryon matrix elements are now calculated under the same framework without introducing new parameters.
		The matrix elements  of  4-quark operators  are  found to be nearly twice larger than the previous ones. The calculated branching fractions of $ \Xi_c ^0 \to \Lambda_c ^+\pi^-$
		and $\Xi_b^- \to \Lambda^0_b \pi^-$ are both in agreement with current experimental
		results. For the yet-to-be-measured heavy-flavor-conserving decays, we find
		${\cal B}( \Xi_c ^+ \to \Lambda_c ^+\pi^0) = (13.8 \pm 1.4)\times 10^{-3}$, ${\cal B}( \Xi_b^0 \to \Lambda^0_b \pi^0) = (2.6 \pm 0.3)\times 10^{-3}$,
		${\cal B}( \Omega_c ^0 \to \Xi_c^+\pi^-) = (2.0 \pm 0.2)\times 10^{-3}$ and ${\cal B}( \Omega_c ^0 \to \Xi_c^0\pi^0)= (1.1 \pm 0.1)\times 10^{-3}$. They are all accessible to LHCb, Belle and Belle II and can be tested in the near future. 
		
	\end{abstract}

	\maketitle
	
	\section{Introduction}
	
	It is known that a rigorous and reliable approach for describing the nonleptonic decays of heavy baryons, especially the nonfactorizable effects, does not exist. Nevertheless, there is a special class of weak decays of heavy baryons that deserves special attention,
	namely, heavy-flavor-conserving (HFC) nonleptonic decays such as the singly
	Cabibbo-suppressed decays $\Xi_Q\to\Lambda_Q\pi$ and $\Omega_Q\to\Xi_Q\pi$ with $Q=c,b$ \cite{Cheng:HFC}.
	They can be studied in a more trustworthy way. Since the pion produced in HFC decays are soft, it is legitimate to apply for the soft-pion theorem to evaluate the nonfactorizable contributions in both $S$- and $P$-wave amplitudes. If the heavy quark behaves a spectator, then we will have another great simplification, namely, the $P$-wave amplitude will vanish in $\Xi_Q\to\Lambda_Q\pi$ decays in the heavy quark limit, while the $S$-wave vanishes in $\Omega_Q\to\Xi_Q\pi$ decays in the limit of $m_Q\to \infty$.
	Indeed, this is the case for $\Xi_b\to\Lambda_b\pi$ and  $\Omega_b\to\Xi_b\pi$ decays as the $b$ quark does not participate in weak interactions in these $b$-flavor-conserving decays. However,  charm-flavor-conserving decays $\Xi_c\to \Lambda_c\pi$ and $\Omega_c\to \Xi_c\pi$ receive additional contributions from the $W$-exchange diagrams via $cs\to dc$ transition. Hence, the charm quark no longer acts as a spectator and the  $P$-wave ($S$-wave) will not diminish in $\Xi_c\to \Lambda_c\pi$ ($\Omega_c\to \Xi_c\pi$) decays in the heavy quark limit.

	In the soft-pion limit, the HFC decay amplitudes can be formulated in terms of the matrix elements of the current (i.e. 2-quark operator) and 4-quark operators~\cite{PreviousWork}.  Previously, they were estimated within the static bag model by two of us~(HYC and FX).
	However, the calculated branching fraction of $\Xi_c^0\to \Lambda_c^+\pi^-$ of order $1.76\times 10^{-3}$~\cite{PreviousWork} is too small by a factor of 3 compared to the recent LHCb measurement of $(5.5\pm0.2\pm1.8)\times 10^{-3}$ \cite{LHCb:HFCc}.
	
	The problem can be traced back to the fact that the description of the bag model is a semi-classical one.
	Quarks are treated as quantum objects as they obey the free Dirac equation, while the boundary of  the bag is a classical object, an infinite potential barrier, possessing a definite position.
	Clearly, a static bag is not invariant under space translation, and it is impossible for a static bag  to be at rest in the quantum mechanics as 3 momenta are the generators of  space translation. 
	The problem is also related to the salient feature that the bag quarks are unentangled. Although the expectation value of the 3-momentum vanishes
	\begin{equation}\label{eq1}
		\langle {\bf p }_q  \rangle = 0\,,
	\end{equation}
	it is impossible for $\langle {\bf p } _q^2 \rangle$ to be zero due to
	\begin{equation}\label{eq2}
		\langle {\bf p }_q ^2 \rangle  = E_q^2  - M^2_q> 0 \,,
	\end{equation}
	where $E_q$, ${\bf p}_q$ and $M_q$ are the energy, 3-momentum and mass, respectively,  of the quark $q$.
	Here, we have used the fact that the bag quarks obey the free Dirac equation and possess a definite energy.   Combining Eqs.~\eqref{eq1} and \eqref{eq2}, 
	we arrive at~\footnote{
		It can also be deduced from the Heisenberg uncertainty principle. As the baryon is localized, it cannot possess a definite 3-momentum. 
	}
	\begin{eqnarray}
		&&\langle {\bf p}_{{\cal B}}  \rangle = 
		\langle  {\bf p}_{q_1 }+{\bf p}_{q_2 }+{\bf p}_{q_3 }   \rangle = 0 \,,
		\nonumber\\
		&&\langle {\bf p}_{{\cal B}}^2 \rangle = 
		\langle \left( {\bf p}_{q_1 }+{\bf p}_{q_2 }+{\bf p}_{q_3 } \right) ^2 \rangle = \langle {\bf p }_{q_1} ^2 \rangle  + \langle {\bf p }_{q_2} ^2 \rangle  + \langle {\bf p }_{q_3}^2 \rangle >0  \,,
	\end{eqnarray}
	where ${\bf p}_{{\cal B}}$ and  ${\bf p}_{q_i}$ are the 3-momenta of the baryon and the $i$-th quarks, respectively, and we have used $\langle {\bf p}_{q_i} \cdot {\bf p}_{q_j}\rangle  = \langle {\bf p}_{q_i}\rangle \cdot \langle {\bf p}_{q_j} \rangle $ for $i\neq j$ as the quarks are unentangled. 
	The variance of the baryon's 3-momentum $\sigma^2_{{\bf p}_{\cal B}}\equiv \langle {\bf p}^2 _{{\cal B}}\rangle-\langle {\bf p} _{{\cal B}}\rangle^2 $ would be referred to the center-of-mass motion~(CMM) of the bag.
	It is straightforward to see that the CMM is Lorentz invariant:
	\begin{eqnarray}
		\sigma^2_{{\bf p}_{\cal B}} &=&
		\langle E_{\cal B}^2 \rangle  - \langle  p^\mu _{{\cal B}} p^\nu _{{\cal B}}   \rangle g_{\mu \nu} - \langle E_{\cal B} \rangle ^2  +  \langle  p ^\mu  _{{\cal B}}\rangle\langle  p^\nu  _{{\cal B}}\rangle g_{\mu \nu }\nonumber\\
		&=&
		- \langle  p^\mu _{{\cal B}} p^\nu _{{\cal B}}   \rangle g_{\mu \nu} +  \langle  p ^\mu  _{{\cal B}}\rangle\langle  p^\nu  _{{\cal B}}\rangle g_{\mu \nu } \,,
	\end{eqnarray}
	where $p^\mu_{{\cal B}}$ is the 4-momentum of the baryon. Since the last two terms in the above equation cancel each other, a physical bag with a definite momentum should not have 
	a CMM as $\sigma^2_{{\bf p}_{\cal B}}=0$.
	However, in the static bag model we have $\sigma^2_{{\bf p}_{\cal B}}> 0$. This unwanted CMM is an issue and it should be removed for a consistent treatment.

	In this update, we shall make an improvement on this shortcoming by getting rid of the CMM of the bag. In particular, it has been shown that  the axial form factors are found to be $15\%$  larger~\cite{Liu:2022pdk} once the CMM issue is remedied. 
	
	The layout of this work is organized as follows.  In Sec.~\MakeUppercase{\romannumeral 2}, we recall the expressions of the HFC decay amplitudes given previously.  In Sec.~\MakeUppercase{\romannumeral 3}, we construct the baryon wave functions free of the CMM effect and sketch the methods of computing the matrix elements. Numerical results and discussions are presented in Sec.~\MakeUppercase{\romannumeral 4}. We conclude our study in Sec.~\MakeUppercase{\romannumeral 5}. Baryon wave functions are summarized in Appendix A. Appendix B is devoted to an explicit evaluation of a given matrix element of 4-quark operators.

	\section{Formalism}
	In this section we recapitulate the formalism outlined in Ref.  \cite{PreviousWork}.
	In the HFC processes, the $W$-exchange through $su \to ud$ transitions is described by the following effective Hamiltonian,
	\begin{eqnarray}
		{\cal H}_{\rm eff}^u  = \frac{G_F}{\sqrt{2}} V_{ud}^* V_{us} \left( c_1O_1^u +c_2 O^u_2 \right) \,,
	\end{eqnarray}
	where $G_F$ is the Fermi constant,  $c_{1,2}$ the Wilson coefficients and $V_{qq'}$  the Cabibbo-Kobayashi-Maskawa (CKM) matrix elements. 
	For HFC decays of charmed baryons, they receive additional $W$-exchange contributions from the $cs \to dc$ transition. The relevant effective Hamiltonian reads
	\begin{eqnarray}
		{\cal H}_{\rm eff}^c  = \frac{G_F}{\sqrt{2}} V_{cd}^* V_{cs} \left( c_1O_1^c +c_2 O^c_2 \right) \,,
	\end{eqnarray}
	with 
	\begin{equation}
		O_1^q = ( d ^\dagger_{\alpha} L^\mu  q _\alpha ) ( q^\dagger_\beta L_\mu s _\beta) \,, ~~~
		O_2^q = ( q ^\dagger_{\alpha} L^\mu  q _\alpha ) ( d^\dagger_\beta L_\mu s _\beta) \,,
	\end{equation}
	where the Greek~$(\alpha,\beta,\gamma)$  alphabets stand for the color indices and  $L^\mu = \gamma^0 \gamma^\mu(1-\gamma_5)$. 
	
	The general amplitude for $\mathcal{B}_{i}   \to \mathcal{B}_{f}  + P  $ is given as 
	\begin{equation}
		M\left(\mathcal{B}_{i} \rightarrow \mathcal{B}_{f}+P\right)=i \bar{u}_{f}\left(A-B \gamma_{5}\right) u_{i}\,,
	\end{equation}
	where $\mathcal{B}_{i(f)}$ are the parent (daughter) baryons with $u_{i(f)}$  their  Dirac spinors, and $A$ and $B$ correspond to the parity-violating (PV) $S$-wave and parity-conserving (PC) $P$-wave amplitudes, respectively. The partial wave amplitudes of HFC $\Xi_Q\to\Lambda_Q\pi$ and $\Omega_Q\to\Xi_Q\pi$ decays  have been studied in Ref. \cite{PreviousWork} for $Q=b,c$.
	The expressions of 
	the $S$-wave amplitudes  are collected as follows~\cite{PreviousWork} \footnote{The reader may notice that the expressions of the $S$- and $P$-wave amplitudes in this work have opposite signs to that given in Ref.  \cite{PreviousWork}. We use the PDG convention $\langle \pi^-(q)|A_\mu|0\rangle=-if_\pi q_\mu$ \cite{pdg}, which was erroneously stated in the first footnote of
		Ref.~\cite{PreviousWork}, and then follow the scenario outlined in Ref. \cite{Zou:2019kzq} for partial-wave amplitudes.}
	\begin{eqnarray}\label{Swave}
		&&A^{ \Xi_{c}^{0} \rightarrow \Lambda_{c}^{+} \pi^{-}}= \zeta \left[a_{1} f_{\pi}^{2}\left(m_{\Xi_{c}}-m_{\Lambda_{c}}\right) f_{1}^{\Lambda_{c}^{+} \Xi_{c}^{0}} -
		c_- \left(a_{\Lambda_c^+\Xi_c^+}^ u  -a_{\Lambda_c^+\Xi_c^+}^ c\right)\right]\,,\nonumber\\
		&&A^{\Xi_{c}^{+} \rightarrow \Lambda_{c}^{+} \pi^{0}}=\frac{\zeta}{\sqrt{2}} \left[a_{2} f_{\pi}^{2}\left(m_{\Xi_{c}}-m_{\Lambda_{c}}\right) f_{1}^{\Lambda_{c}^{+} \Xi_{c}^{+}}-c_- \left(a_{\Lambda_c^+\Xi_c^+}^ u  -a_{\Lambda_c^+\Xi_c^+}^ c\right)\right] \,, \nonumber\\
		&&A^{ \Xi_{b}^{-} \rightarrow \Lambda_{b}^{0} \pi^{-}}= \zeta \left[a_{1} f_{\pi}^{2}\left(m_{\Xi_{b}}-m_{\Lambda_{b}}\right) f_{1}^{\Lambda_{b}^{0} \Xi_{b}^{-}} -
		c_- a_{\Lambda_b^0\Xi_b^0}^ u  \right]\,,\nonumber\\
		&&A^{\Xi_{b}^{0} \rightarrow \Lambda_{b}^{0} \pi^{0}}=\frac{\zeta}{\sqrt{2}} \left[a_{2} f_{\pi}^{2}\left(m_{\Xi_{b}}-m_{\Lambda_{b}}\right) f_{1}^{\Lambda_{b}^{0} \Xi_{b}^{0}}-c_- a_{\Lambda_b^0\Xi_b^0}^ u   \right] \,, \nonumber\\
		&& A^{\Omega_{c}^{0} \rightarrow \Xi_{c}^{+} \pi^{-}}= -\zeta  
		c_- a_{\Xi_c^0 \Omega_c^0}^c 
		\,,
		\nonumber\\
		&& A^{\Omega_{c}^{0} \rightarrow \Xi_{c}^{0} \pi^{0}}= \frac{\zeta}{\sqrt{2}}
		c_-  a_{\Xi_c^0 \Omega_c^0}^c 
		\,,\nonumber\\
		&&A^{ \Omega_{b}^{-} \rightarrow \Xi_{b}^{0} \pi^{-}}=A^{\Omega_{b}^{0} \rightarrow \Xi_{b}^{0} \pi^{0}}=0\,,
	\end{eqnarray}
	whereas the $P$-wave amplitudes read
	\begin{eqnarray}\label{Pwave}
		&&
		B^{\Xi_{c}^{0} \rightarrow \Lambda_{c}^{+} \pi^{-}} = 
		\zeta c_-\left(g_{\Lambda_{c}^{+} \Sigma_{c}^{0}}^{A\left(\pi^{-}\right)} \frac{m_{\Lambda_{c}^{+}}+m_{\Sigma_{c}^{0}}}{m_{\Xi_{c}^{0}}-m_{\Sigma_{c}^{0}}} a^c_{\Sigma_{c}^{0} \Xi_{c}^{0}}+ a^c_{\Lambda_{c}^{+} \Xi_{c}^{\prime+}} \frac{m_{\Xi_{c}^{0}}+m_{\Xi^{\prime+}_c}}{m_{\Lambda_{c}^{+}}-m_{\Xi_{c}^{+}}} g_{\Xi_{c}^{\prime+} \Xi_{c}^{0}}^{A\left(\pi^{-}\right)}\right),\nonumber\\
		&&
		B^{\Xi_{c}^{+} \rightarrow \Lambda_{c}^{+} \pi^{0}} = 
		\sqrt{2} \zeta c_-\left(g_{\Lambda_{c}^{+} \Sigma_{c}^{+}}^{A\left(\pi^{0}\right)} \frac{m_{\Lambda_{c}^{+}}+m_{\Sigma_{c}^{+}}}{m_{\Xi_{c}^{+}}-m_{\Sigma_{c}^{+}}} a^c_{\Sigma_{c}^{+} \Xi_{c}^{+}}+ a^c_{\Lambda_{c}^{+} \Xi_{c}^{\prime+}} \frac{m_{\Xi_{c}^{+}}+m_{\Xi_{c}^{\prime+}}}{m_{\Lambda_{c}^{+}}-m_{\Xi_{c}^{\prime+}}} g_{\Xi_{c}^{\prime+} \Xi_{c}^{+}}^{A\left(\pi^{0}\right)} \right) \,,\nonumber\\
		&&
		B^{ \Xi_{b}^{-} \rightarrow \Lambda_{b}^{0} \pi^{-}}=B^{\Xi_{b}^{0} \rightarrow \Lambda_{b}^{0} \pi^{0}}=0\,, \nonumber \\
		&&
		B^{\Omega_{c}^{0} \rightarrow \Xi_{c}^{+} \pi^{-}}= -\zeta 
		\left[  
		a_{1} f_{\pi}^2\left(m_{\Omega_{c}}+m_{\Xi_{c}}\right) g_{1}^{\Xi_{c}^{+} \Omega_{c}^{0}}  + c_-
		g^{A(\pi^-)} _{ \Xi_c^+\Xi_c^{\prime 0 }}  \frac{m_{\Xi_c^+} + m_{\Xi_c^{\prime 0 }} }{m_{\Omega_c^0}     - m_{ \Xi_c^{\prime 0 }  }}      a_{\Xi_c^{\prime 0}  \Omega_c^0}^c  
		\right] \,,\nonumber\\
		&&B^{\Omega_{c}^{0} \rightarrow \Xi_{c}^{0} \pi^{0}}= - \frac{\zeta}{\sqrt{2}}
		\left[  
		a_{2} f_{\pi}^2\left(m_{\Omega_{c}}+ m_{\Xi_{c}}\right) g_{1}^{\Xi_{c}^{0} \Omega_{c}^{0}} + 
		2 c_- g^{A(\pi^0)} _{ \Xi_c^0\Xi_c^{\prime 0 }}  \frac{m_{\Xi_c^+} + m_{\Xi_c^{\prime 0 }} }{m_{\Omega_c^0}     - m_{ \Xi_c^{\prime 0 }  }}      a_{\Xi_c^{\prime 0}  \Omega_c^0}^c  
		\right] \,,\nonumber\\
		&&
		B^{\Omega_b ^ - \to \Xi_b^ 0 \pi^- } = - \zeta a_1 f_\pi^2 ( m_{\Omega_b} + m_{\Xi_b}  ) 
		g_1^{\Xi_b^0 \Omega_b^-} \,,
		\nonumber\\
		&&
		B^{\Omega_b ^ - \to \Xi_b^ - \pi^0 } = - \frac{\zeta}{\sqrt{2}} a_2 f_\pi^2 ( m_{\Omega_b} + m_{\Xi_b}  ) 
		g_1^{\Xi_b^- \Omega_b^-} \,,
	\end{eqnarray}
	along with the following abbreviations
	\begin{eqnarray}
		\zeta \equiv \frac{G_F}{\sqrt{2} f_\pi } V_{ud}^* V_ {us}\,, ~~~c_- \equiv  \frac{1}{2}\left( c_1 - c_2 \right) \,. 
	\end{eqnarray}
	
	In Eqs. (\ref{Swave}) and (\ref{Pwave}), the factorizable contributions are expressed in terms of parameters $a_{1,2}$ given by 
	\begin{eqnarray} \label{a1a2}
		a_1=c_1+{c_2\over N_c}, \qquad a_2=c_2+{c_1\over N_c},
	\end{eqnarray}
	and the form factors $f_1^{{\cal B}'{\cal B}}$ and $g^{{\cal B}'{\cal B}}_1$  defined by
	\begin{eqnarray}\label{form}
		&& \langle {\cal B}' | \overline{q'}_ \alpha  (0) \gamma ^\mu q_\alpha(0)| {\cal B}\rangle 
		=\overline{u}_{{\cal B}'} \left(
		f^{{\cal B}'{\cal B}}_1(q^2) \gamma^\mu - f^{{\cal B}'{\cal B}}_2 (q^2)i \sigma^{\mu \nu} \frac{q_\nu}{ m_{{\cal B}} }   +f^{{\cal B}'{\cal B}}_3(q^2) \frac{q^\mu}{m_{{\cal B}}}
		\right)u_{{\cal B}}\,,\\
		&&\langle {\cal B}' | \overline{q'}_ \alpha  (0) \gamma ^\mu\gamma_5  q_\alpha(0)| {\cal B}\rangle 
		=\overline{u}_{{\cal B}'} \left(
		g_1^{{\cal B}'{\cal B}}(q^2) \gamma^\mu - g_2^{{\cal B}'{\cal B}} (q^2)i \sigma^{\mu \nu} \frac{q_\nu}{ m_{{\cal B}} }   +g_3^{{\cal B}'{\cal B}}(q^2) \frac{q^\mu}{m_{{\cal B}}}
		\right)\gamma_5 u_{{\cal B}}\,,\nonumber
	\end{eqnarray}
	where  $(q',q) = (u,s)$ and $(d,s)$, depending on the isospin of ${\cal B}'$ and ${\cal B}$. 
	In Eqs. (\ref{Swave}) and (\ref{Pwave}), the form factors $f^{{\cal B}'{\cal B}}_1$ and $g^{{\cal B}'{\cal B}}_1$ are evaluated at $q^2=m_\pi^2$.
	Note that the Wilson coefficients $c_{1,2}$ and the parameters $a_{1,2}$ are evaluated at the scale  $\mu=m_c$ and $\bar m(m_b)$
	for charmed and bottom baryons, respectively. We follow Ref.~\cite{PreviousWork} to employ
	the Wilson coefficients $c_1=1.336$ and $c_2=-0.621$ at the scale $\mu=m_c$ 
	and $c_1=1.139$ and $c_2=-0.307$ at $\mu=\bar m_b(m_b)$. 
	
	Nonfactorizable contributions arising from the  $W$-exchange diagrams induced by $su\to ud$ and $cs\to dc$ transitions are dedicated by the PC matrix elements $a^{u}_{{\cal B} ' {\cal B}  }$ and $a^c_{{\cal B} ' {\cal B}  }$, respectively, of  4-quark operators
	defined by \footnote{Recall that the commutator term of the $S$-wave in the soft-pion limit is proportional to
		$\langle {\cal B}_f | [Q_5^a, {\cal H}_{\rm eff}^{\rm pv}]|{\cal B}_i \rangle=- \langle {\cal B}_f |[Q^a, {\cal H}_{\rm eff}^{\rm pc}]|{\cal B}_i\rangle$.}
	\begin{equation} \label{12}
		\langle {\cal B}' | O_1^{u(c)}-O_2^{u(c)}   | {\cal B}  \rangle = \overline{u}_{{\cal B}'} \left(
		a^{u (c)}_{{\cal B} ' {\cal B}  }  + b^{u(c)} _{{\cal B} ' {\cal B}  } \gamma_5 
		\right)u_{{\cal B} }\,,
	\end{equation}
	where  both $a^{u (c)}_{{\cal B} ' {\cal B}  }  $ and $b^{u (c)}_{{\cal B} ' {\cal B}  }  $ have canonical dimension 3. 
	The axial couplings $g^{A(\pi^-)}_{{\cal B}' {\cal B}}$ and  $g^{A(\pi^0)}_{{\cal B}' {\cal B}}$ in Eq. (\ref{Pwave}) are  the special cases of $g_1$,  obtained by taking $(q',q) = (u ,d)$  and $(q',q) = ((u,u)-(d,d))/2$ at $q^2=0$, respectively. 
	
	From Eqs. (\ref{Swave}) and (\ref{Pwave}) we see that the $S$-wave of $\Omega_b^-\to \Xi_b \pi$ and the $P$-wave of $\Xi_b \to \Lambda_b \pi$ vanish. This is ascribed to the fact that the $b$ quark in these HFC decays does not participate in weak interactions. The diquark transition $1^+\to 0^+ + 0^-$ in  $\Omega_b^-\to \Xi_b \pi$ in $S$-wave is prohibited by conservation of angular momentum. Likewise, the diquark transition $0^+\to 0^+ + 0^-$ in  $\Xi_b \to \Lambda_b \pi$ in $P$-wave is not allowed by the same token.

	\section{model calculation}	
	In this section we shall calculate the matrix elements of the current and 4-quark operators using the homogeneous bag model in which the unwanted CMM of the bag is removed~\cite{Liu:2022pdk}.

	\subsection{Baryon wave functions}
	The baryon wave functions are given in Appendix A. Here we take the antitriplet charmed baryons for an illustration 
	\begin{eqnarray}
		&&|\Xi_c^0 , \uparrow \rangle = \int\frac{1}{\sqrt{6} } \epsilon^{\alpha \beta \gamma} d_{a\alpha}^\dagger(\vec{x}_1) s_{b\beta}^\dagger(\vec{x}_2) c_{c\gamma}^\dagger (\vec{x}_3) \Psi^{abc}_{{A_\uparrow(dsc)}} (\vec{x}_1,\vec{x}_2,\vec{x}_3) [d^3  \vec{x}] | 0\rangle\,,\nonumber\\
		&&|\Xi_c^+ , \uparrow \rangle = \int\frac{1}{\sqrt{6} } \epsilon^{\alpha \beta \gamma} u _{a\alpha}^\dagger(\vec{x_1}) s_{b\beta}^\dagger(\vec{x_2}) c_{c\gamma}^\dagger (\vec{x_3}) \Psi^{abc}_{{A_\uparrow(usc)}} (\vec{x}_1,\vec{x}_2,\vec{x}_3) [d^3  \vec{x}] | 0\rangle\,,\nonumber\\
		&&	|\Lambda_c^+, \uparrow  \rangle = \int\frac{1}{\sqrt{6} } \epsilon^{\alpha \beta \gamma} d _{a\alpha}^\dagger(\vec{x}_1) u_{b\beta}^\dagger(\vec{x}_2) c_{c\gamma}^\dagger (\vec{x}_3) \Psi_{A_\uparrow(duc)}^{abc} (\vec{x}_1,\vec{x}_2,\vec{x}_3) [d^3  \vec{x}] | 0\rangle\,,
	\end{eqnarray}
	where $[d^3 \vec{x}]= d^3\vec{x}_1 d^3\vec{x}_2 d^3\vec{x}_3$, the Latin~$(a,b,c)$ alphabets are  the spinor indices, and  $\epsilon$ is a totally anti-symmetric tensor. 
	Compared to  the static bag wave functions given in  Ref.~\cite{Singly Cabibbo}, 
	the wave functions here are expressed in terms of creation operators and the Fermi statistics is taken care of by the anti-commutation relation 
	\begin{equation}\label{anti}
		\{ q _{a \alpha} (\vec{x}) , q ^\dagger _{b\beta} ( \vec{x}')  \} = \delta_{ab } \delta_{\alpha \beta}\delta^3(\vec{x} -\vec{x}') \,. 
	\end{equation}
	For the spatial distribution $\Psi$ of the quarks, we shall describe it using  the homogeneous bag model first proposed in Ref. \cite{bagO}
	\begin{equation} \label{xdelta}
		\begin{aligned}
			\Psi_{A_{\updownarrow}(q_1q_2q_3)}^{abc} ( \vec{x}_1, \vec{x}_2 ,  \vec{x}_3 ) =& \frac{{\cal N}}{\sqrt{2}} \int \left [ \phi^a_{q_1\uparrow}(\vec{x}_1- \vec{x}_\Delta) \phi^b_{q_2\downarrow}(\vec{x}_2-\vec{ x}_\Delta) \right. \\
			&\left. - \phi^a_{q_1\downarrow}(\vec{x}_1- \vec{x}_\Delta) \phi^b_{q_2\uparrow}(\vec{x}_2-\vec{ x}_\Delta) \right]  \phi^c_{q_3\updownarrow}(\vec{x}_3- \vec{x}_\Delta) d^3 \vec{x}_\Delta,
		\end{aligned}
	\end{equation}
	where ${\cal N}$ is a normalization constant and 
	$\phi^a_q$ are the bag wave functions in the static  bag model, 
	\begin{equation}\label{quark_wave_function}
		\phi_{q\updownarrow} (\vec{r}) = \left(
		\begin{array}{c}
			u_c(r)   \chi_\updownarrow\\
			v_c(r) \hat{r} \cdot \vec{\sigma} \chi_\updownarrow\\
		\end{array}
		\right)\equiv 
		\left\{
		\begin{array}{ll}
			\left(
			\begin{array}{l}
				E_+ j_0 (|{\bf p}_q|r )   \chi_\updownarrow\\
				E_-  j_1 (|{\bf p}_q|r ) \hat{r} \cdot \vec{\sigma} \chi_\updownarrow\\
			\end{array}
			\right)\,& ~~~\text{for} ~~r \leq R\,, \\
			0&~~~\text{for} ~~r > R\,,
		\end{array}
		\right.
	\end{equation}
	with $\chi _\uparrow = (1,0)^T$ and $\chi_\downarrow = (0,1)^T$ representing $J_z = \pm 1/2$, $E_\pm = \sqrt{E_q\pm M_q} $, and $j_{0,1}$ the spherical Bessel functions.
	Note that the spin structure of Eq.~\eqref{xdelta} is simply 
	\begin{equation}
		\frac{1}{\sqrt{2}}\left( \uparrow\downarrow-\downarrow\uparrow 
		\right)\updownarrow\,,
	\end{equation}
	and the heavy quark clearly points to the same direction with the antitriplet baryon.

	The integral of $d^3\vec{x}_\Delta$ in Eq.~\eqref{xdelta} plays an essential role of removing the CMM, rendering the wave function to be invariant under  space translation.
	To see this, we apply  the space translation operator ${\cal T}_{\vec{d}}$ on $|\Lambda_c^+\rangle$ for example,
	\begin{eqnarray}
		&&{\cal T}_{\vec{d}}\, |\Lambda_c^+\rangle  = 
		\int\frac{1}{\sqrt{6} } \epsilon^{\alpha \beta \gamma} d _{a\alpha}^\dagger(\vec{x}_1+\vec{d}\,) u_{b\beta}^\dagger(\vec{x}_2+\vec{d}\,) c_{c\gamma}^\dagger (\vec{x}_3+\vec{d}\,) \Psi_{A_\uparrow(duc)}^{abc} (\vec{x}_1,\vec{x}_2,\vec{x}_3) [d^3  \vec{x}] | 0\rangle\,,\nonumber\\
		&&
		=  \int\frac{1}{\sqrt{6} } \epsilon^{\alpha \beta \gamma} d _{a\alpha}^\dagger(\vec{x}_1) u_{b\beta}^\dagger(\vec{x}_2) c_{c\gamma}^\dagger (\vec{x}_3) \Psi_{A_\uparrow(duc)}^{abc} (\vec{x}_1-\vec{d},\vec{x}_2-\vec{d},\vec{x}_3-\vec{d}\,) [d^3  \vec{x}] | 0\rangle\,,\nonumber\\
		&&
		=  \int\frac{1}{\sqrt{6} } \epsilon^{\alpha \beta \gamma} d _{a\alpha}^\dagger(\vec{x}_1) u_{b\beta}^\dagger(\vec{x}_2) c_{c\gamma}^\dagger (\vec{x}_3) \Psi_{A_\uparrow(duc)}^{abc} (\vec{x}_1,\vec{x}_2,\vec{x}_3) [d^3  \vec{x}] | 0\rangle = | \Lambda_c^+\rangle\,,
	\end{eqnarray}
	where use of 
	${\cal T}_{\vec{d}}\,q_{a\alpha}^\dagger(\vec{x})  {\cal T}_{\vec{d}}^\dagger  = q_{a\alpha}^\dagger(\vec{x}+\vec{d})$ and 
	${\cal T}_{\vec{d}}\, |0\rangle = |0\rangle$ has been made in the first line of the equation, and the second and third lines are obtained by changing the integration variables 
	\begin{eqnarray}
		\vec{x}_{1,2,3} \to \vec{x}_{1,2,3} - \vec{d}\,,
	\end{eqnarray}
	and 
	\begin{eqnarray}
		\vec{x}_{\Delta} \to \vec{x}_{\Delta} + \vec{d}\,,
	\end{eqnarray}
	respectively.
	As the 3-momentum operator ${\bf {\hat P}} $ is the generator of space translation, we deduce that 
	\begin{equation}
		{\bf {\hat P}} | \Lambda_c^+\rangle = {\bf {\hat P}} ^2 | \Lambda_c^+\rangle = 0\,.
	\end{equation}
	Thus, the CMM is taken away from the static bag.
	Throughout this work, we give the integration variables $\vec{x}_\Delta$ an additional meaning.
	By eliminating every $d^3 \vec{x}_\Delta$ integrals,  the formulae  would be reduced to the ones in  the static limit~\cite{PreviousWork}.

	For a baryon at rest (i.e. with zero momentum), quarks distribute homogeneously all over the
	space to fulfill the translation-invariant requirement.
	Now the baryons are described by a linear superposition of infinite bags.  
	Although quarks are no longer localized, they are confined with each other in the sense that 
	\begin{equation}\label{neq}
		\Psi_{A_{\updownarrow}(q_1q_2q_3)}^{abc} ( \vec{x}_1, \vec{x}_2 ,  \vec{x}_3 ) = 0\,,~~~\text{for}~~|\vec{x}_i - \vec{x}_j| >2R\,,
	\end{equation}
	with $i,j = 1,2,3$ and $i\neq j$. It can be proved by noticing that the integrand in Eq.~\eqref{xdelta} vanishes when $|\vec{x}_\Delta - \vec{x}_i | > R$. While Eq~\eqref{neq} holds in the homogeneous bag model because the locations of the quarks are entangled, it is also valid in the static bag model since quarks are localized.


	Of course, it is possible to construct other baryon wave functions free of the CMM issue in different bag models. Nevertheless, we believe that the homogeneous bag model for $\Psi$ which we have adopted here is the simplest one as it requires no new parameter and the static limit can be easily recovered.
	
	By taking the normalizing condition as 
	\begin{equation}
		\langle {\bf p}_{{\cal B}} \,| {\bf p}_{{\cal B}}' \rangle  = u^\dagger  {u'} (2\pi)^3 \delta^3({\bf p}_{{\cal B}} - {\bf p}_{{\cal B}}' ) \,,
	\end{equation}
	we find~\cite{Liu:2022pdk}
	\begin{equation}\label{normal}
		\frac{\overline{u} u }{{\cal N}^2 }= \int d^3\vec{x}_\Delta \prod_{i=1,2,3}\int    \phi_{q_i}^\dagger \left(\vec{x}_i^+\right) \phi_{q_i} \left(\vec{x}_i ^-\right) d^3\vec{x}_i\,,
	\end{equation}
	where $\vec{x}^\pm_i = \vec{x}_i \pm \vec{x}_\Delta /2 $.
	In Eq.~\eqref{normal}, the spin direction of $\phi_{q_i}$ is not important as it leads to the same result. In the following, we will suppress the spin indices whenever they make no difference.

	To cooperate with the soft pion limit, we take the 3-momenta of the initial and final baryons to be zero for consistency. 
	In particular, the matrix elements of 4-quark operators would simply vanish if the baryons have different momenta as ${\cal H}_{\rm eff}$ respects the energy-momentum conservation law.
	Since $q^2 =m_\pi^2\approx (M_{\Xi_Q} - M_{\Lambda_Q})^2 \approx (M_{\Omega_Q} - M_{\Xi_Q})^2 \approx 0$, the treatment we adopt here is in fact an excellent approximation.
	For example, the errors of the form factors are of order $10^{-3}$.
	The method of boosting the baryon wave functions requires several additional steps, and the interested readers are referred to Ref.~\cite{Liu:2022pdk} for detail.

	\subsection{Current operators}
	
	To illustrate the  calculation of form factors, we take $\Xi_c^0 \to \Lambda_c^+$ as an example.
	It follows from Eq.~\eqref{anti} that 
	\begin{eqnarray}\label{Lm}
		&&\langle \Lambda_c^+ | u^\dagger_\alpha  (0) L^\mu s_\alpha(0) | \Xi_c^0\rangle  = \int \Psi_{A(duc)}^{ab'c\dagger} (\vec{x}_1,0 , \vec{x}_3) ( L ^\mu)^{b'b}   
		\Psi_{A(dsc)}^{abc}(\vec{x}_1,0, \vec{x}_3)  d^3\vec{x}_1 d^3\vec{x}_3 \\
		&& =\int d^3\vec{y} d^3\vec{y}\,'d^3\vec{x}_1d^3\vec{x}_3 
		\phi_{d}^\dagger \left( \vec{x}_1-\vec{y} \, \right) \phi_{d} \left(\vec{x}_1-\vec{y}\,'  \right) 
		\phi _{u} ^\dagger\left(  -\vec{y} \right) L ^\mu\phi_{s}\left(  -\vec{y}\,' \right) 
		\phi_{c}^\dagger \left(\vec{x}_3   -\vec{y}\right) \phi_{c} \left(\vec{x}_3   -\vec{y}\,' \right) ,\nonumber
	\end{eqnarray}
	With Eq.~\eqref{normal}, it is easily seen that form factors are independent of the normalization condition of the Dirac spinors~$(\overline{u}u)$.
	By changing the integration variables
	\begin{equation}
		\left(\vec{y} , \vec{y}\,', \vec{x}_1
		,\vec{x}_3
		\right) \to \left(
		-\vec{x}_2 -\frac{1}{2}\vec{x}_\Delta , -\vec{x}_2 + \frac{1}{2}\vec{x}_\Delta , \vec{x}_1 - \vec{x}_2, \vec{x}_3 - \vec{x}_2
		\right)
	\end{equation}
	we recast Eq.~\eqref{Lm} to  a more symmetric form
	\begin{equation}\label{master}
		\begin{aligned}
			\langle \Lambda_c^+ | u^\dagger_\alpha  (0) L^\mu s_\alpha(0) | \Xi_c^0\rangle   = 	\int  [d^3\vec{x}]  d^3\vec{x}_\Delta & \phi _{u} ^\dagger\left(\vec{x}_2 ^+ \right) L ^\mu\phi_{s}\left(\vec{x}_2 ^-  \right) \\ 
			& \times \phi_{d}^\dagger \left(\vec{x}_1 ^+ \right) \phi_{d} \left(\vec{x}_1 ^-  \right) \phi_{c}^\dagger \left(\vec{x}_3 ^+ \right) \phi_{c} \left(\vec{x}_3 ^-  \right) ,
		\end{aligned} 
	\end{equation}
	where
	the static limit is recovered by taking $\vec{x}_i^\pm \to \vec{x}_i$. 
	By matching Eqs.~\eqref{form} and \eqref{master}, we find 
	\begin{equation}
		(f_1 \,,~g_1)=( 0.978\,,~0) \,,
	\end{equation}
	where we have used~\cite{DeGrand:1975cf}
	\begin{equation}\label{input}
		R = ( 5.0 \pm 0.1)~\text{GeV}^{-1}\,,~~~M_u=0\,,~~~
		M_s = 0.28 ~\text{GeV} \,,~~~M_c = 1.655~\text{GeV} \,.
	\end{equation}
	For form factors, the uncertainties of the bag radius can be neglected for the precision under consideration in this work. 
	Notice that  $f_1$  is very close to 0.985 obtained  from the MIT bag model~\cite{PreviousWork}. 
	It is ascribed to that $f_1= 1 $ is protected by the  $SU(3)_F$ symmetry and 
	hence a large correction is not expected. 
	
	Since the $c$ quark shares the same spin direction with $\Xi_c^0$ and $\Lambda_c^+$, we are forced to choose both baryon spins aligned in the same direction to get a non-vanishing  result.  Thus,  the spin flipping term $g_{1}$  vanishes, providing that the spin of the antitriplet baryon arises solely from the $c$ quark.   It can also be understood as the transition of the diquark $0^+ \to 0^+ +\pi$ is forbidden by conservation of angular momentum and parity~\cite{Cheng:HFC, PreviousWork}. Thus, the strong coupling of ${\cal B}_{c}{\cal B}_c \pi$ vanishes and so does $g^\pi_{{\cal B}_{c}{\cal B}_{c}}$.
	On the other hand, the $c$ quark in the sextet baryon does not necessarily share the same spin direction with the baryon and hence the aforementioned argument cannot be applied. 
	
	The relevant form factors are given in Table~\ref{table1}.
	Notice that in contrast to  $f_1$, $g_1$ not protected by the flavor symmetry  and  found to be 16$\%$ larger than the one in the static limit~\cite{PreviousWork}.

	\begin{table}[t]
		
		\caption{ Numerical results of the form factors with $\pi \in \{ \pi^-, \pi^0\}$, depending on the isospin of the baryons. }\label{table1}
		\vskip 0.2cm
		\begin{tabular}{lcccc|ccccc}
			\hline
			\hline
			$( {\cal B}, {\cal B}' ) $	& $( \Lambda^+_c, \Sigma_c^0 ) $ 	& $( \Xi_c^{\prime+}, \Xi_c^{ 0}  ) $&  $( \Xi_c^{+}, \Xi_c^{ \prime +  }  )  $ & $( \Xi_c^{0 }, \Xi_c^{ \prime 0  }  )  $~ &		$( {\cal B}, {\cal B}' ) $& $( \Xi_c^{ +(0) }, \Omega_c^0  ) $& $( \Xi_b^{ 0 (-)  }, \Omega_b^-  ) $\\
			\hline
			$g_{{\cal B}{\cal B}'}^\pi $ &$0.614$&$-0.436$& $-0.218 $ & $0.218 $ & ~~$g_1^{s\to u(d)}  $& $-0.616  $&  $ -0.597$  \\
			\hline
			\hline		
		\end{tabular}
	\end{table}
	\subsection{4-quark operators}
	
	The calculations of  4-quark operators are similar to those of the current ones. 
	We take the $su \to ud$ transition in $\Xi_c^+\to \Lambda_c^+$  to illustrate the method.
	Applying the anti-commutation relation in Eq.~\eqref{anti}, we find 
	\begin{equation}\label{11}
		\begin{aligned}
			\langle \Lambda_c ^+ |& (d^\dagger_\alpha (0)  L^ \mu u _\alpha (0 )  ) ( u^\dagger _\beta(0)   L_\mu  s_\beta(0)  )  | \Xi_c^+\rangle  \\
			&	=\int \Psi_{A(duc)}^{a'b'c\dagger }(0,0,\vec{x}_3) (L  ^\mu) ^{a'a} (L  _\mu) ^ {b'b} \Psi^{abc}_{A(usc)} (0,0,\vec{x}_3) d^3 \vec{x}_3 \,,
		\end{aligned} 
	\end{equation}
	for  $O^u _1$, and 
	\begin{equation}\label{O4mas}
		\begin{aligned}
			\langle \Lambda_c ^+ |& (u^\dagger_\beta (0)  L^ \mu u _\beta (0 )  ) ( d^\dagger _\alpha(0)   L_\mu  s_\alpha (0)  )  | \Xi_c^+\rangle \\
			&	=\int \Psi_{A(duc)}^{a'b'c\dagger }(0,0,\vec{x}_3) (L  ^\mu) ^{b'a} (L  _\mu) ^ {a'b} \Psi^{abc}_{A(usc)} (0,0,\vec{x}_3) d^3 \vec{x}_3 \,,
		\end{aligned} 
	\end{equation}
	for  $O_2^u$.
	With the Fierz identity $(L  ^\mu) ^{b'a} (L  _\mu) ^ {a'b} = - (L  ^\mu) ^{a'a} (L  _\mu) ^ {b'b}$, it is straightforward to show that $\langle \Lambda_c^+ | (O^u_1+ O^u_2) |  \Xi_c^0\rangle=0$, as it should be because the operator $O_1^u+O_2^u$ is symmetric in color indices, while the baryon wave function is color anti-symmetric. This is known
	as a special case of  the  K\"oner-Pati-Woo theorem~\cite{Korner:1970xq}. 
	On the other hand, for the sextet baryon ${\cal B}_6$ we have 
	\begin{equation}
		\langle {\cal B}_{6} |O_1^u | {\cal B}_6 \rangle = 0 \,,
	\end{equation}
	which can be deduced by substituting $S$ for $A$ in Eqs.~\eqref{11} and using  $\Psi_S^{abc} = \Psi_S^{bac}$ along with the Fierz identity.

	Substituting Eq.~\eqref{xdelta} into Eq.~\eqref{11} leads to 
	\begin{eqnarray}\label{O4Ex}
		&&\begin{aligned}
			\sum_{[\lambda]}  \frac{1}{2}	\left(
			-1
			\right)^{\lambda_1- \lambda_3} & \int 
			d^3\vec{x}_3   d^3\vec{y}d^3\vec{y}\,'  \phi_{c}^\dagger \left(\vec{x}_3 - \vec{y} \right) \phi_{c} \left(\vec{x}_3 - \vec{y}\,' \right)\nonumber
			\\
			&\times \phi _{d\lambda_4} ^\dagger\left(-\vec{y} \right) L 
			_\mu\phi_{u\lambda_2}\left(-\vec{y}\,' \right) \phi _{u\lambda_3} ^\dagger\left(-\vec{y} \right) L ^\mu\phi_{s\lambda_1}\left(-\vec{y} \,' \right) \,, 
		\end{aligned} 
		\nonumber \\
		&&\begin{aligned}
			=\sum_{[\lambda]}  \frac{1}{2}	\left(
			-1
			\right)^{\lambda_1- \lambda_3} & \int   d^3\vec{x}_\Delta 
			d^3\vec{x}_3  \phi_{c}^\dagger \left(\vec{x}_3^+ \right) \phi_{c} \left(\vec{x}_3 ^-  \right)
			\\
			&\times \int  d^3\vec{x}  \phi _{d\lambda_4} ^\dagger\left(\vec{x}^ + \right) L 
			_\mu\phi_{u\lambda_2}\left(\vec{x}^- \right) \phi _{u\lambda_3} ^\dagger\left(\vec{x} ^+ \right) L ^\mu\phi_{s\lambda_1}\left(\vec{x} ^- \right) \,, 
		\end{aligned} \nonumber\\
		&&\equiv  \int d^3 \vec{x}_\Delta  {\cal D}_c(\vec{x}_\Delta) \Gamma (\vec{x}_\Delta)
	\end{eqnarray}
	where $[\lambda]$ stands for the possible configurations of spins 
	\begin{equation}
		(\lambda_1,\lambda_2,\lambda_3,\lambda_4) \in \left\{  ( \downarrow, \uparrow, \downarrow, \uparrow) ,  ( \uparrow, \downarrow, \downarrow, \uparrow) ,  (  \uparrow, \downarrow, \uparrow,\downarrow ) , ( \downarrow, \uparrow,\uparrow,  \downarrow) \right\} \,.
	\end{equation}
	To sort out the expression, we have changed the integration variables 
	\begin{eqnarray}
		(\vec{y}, \vec{y}\,',\vec{x}_3) \to \left(-\vec{x}- \frac{1}{2}\vec{x}_\Delta , - \vec{x} + \frac{1}{2} \vec{x}_\Delta, \vec{x}_3  - \vec{x}\right)\,,
	\end{eqnarray}
	and taken the short hand as
	\begin{eqnarray}
		&&{\cal D}_c (\vec{x}_\Delta)  = \int d^3 \vec{x}_3 \phi_c^\dagger (\vec{x}_3^+ )\phi_c (\vec{x}_3^-) \nonumber\\
		&&\Gamma (\vec{x}_\Delta) = 
		\sum_{[\lambda]}  \frac{1}{2}	\left(
		-1
		\right)^{\lambda_1- \lambda_3}
		\int  d^3\vec{x}  \phi _{d\lambda_4} ^\dagger\left(\vec{x}^ + \right) L 
		_\mu\phi_{u\lambda_2}\left(\vec{x}^- \right) \phi _{u\lambda_3} ^\dagger\left(\vec{x} ^+ \right) L ^\mu\phi_{s\lambda_1}\left(\vec{x} ^- \right)\,.
	\end{eqnarray}
	Although Eq.~\eqref{O4Ex} seems complicated, it can be understood as follows:
	\begin{itemize}
		\item The $\vec{x}_{\Delta}$ is  the distance between the static bags of  the initial 
		and final baryons. 
		The integrals of $d^3 \vec{x}_\Delta$ can be viewed as computing the overlapping of the bags.
		\item  The ${\cal D}_c(\vec{x}_\Delta) $ is  the overlapping of the spectator quark in the static bags separated at a distance of $\vec{x}_\Delta$. 
		\item The $\Gamma(\vec{x}_\Delta)$ describes the weak transition, where
		the $(su)$ and $(ud)$ pairs are annihilated and created in the static bags centering at $-\frac{1}{2}\vec{x}_\Delta$ and  $\frac{1}{2}\vec{x}_\Delta$, respectively. 
	\end{itemize}

	To extract  $a^u_{\Lambda_c^+ \Xi_c^+ }$, we choose the parity-conserving part of the operators,  namely 
	\begin{equation}\label{14}
		(L^\mu) ^{ab} (L_\mu)^{cd} \to (V^\mu) ^{ab} (V_\mu)^{cd}+(A^\mu) ^{ab} (A_\mu)^{cd}\,.
	\end{equation}
	Plugging Eq.~\eqref{14}  to Eq.~\eqref{O4Ex} and comparing with Eq.~\eqref{12}, we find 
	\begin{equation}\label{avalue}
		a^u_{\Lambda_c^+ \Xi_c^+} = (3.74 \pm 0.20)\times 10^{-2}~\text{GeV}^3\,,
	\end{equation}
	where the uncertainties arise from the bag radius in Eq.~\eqref{input} and the details are sketched in Appendix B. 
	Note that this value is  consistent with the estimation
	\begin{equation}
		\frac{1}{c_{1}-c_{2}}(5.6 \pm 1.1)\times 10^{-2}\, \mathrm{GeV}^{3}\,,
	\end{equation}
	made in the diquark model~\cite{Diquark}.
	
	In the previous bag model calculation, it was found $a^u_{\Lambda_c^+ \Xi_c^+}=1.67\times 10^{-2}\, {\rm GeV{^3}}$ ~\cite{PreviousWork}.\footnote{The matrix elements $a^u_{\Lambda_c^+ \Xi_c^+}$ and $a^c_{\Lambda_c^+ \Xi_c^+}$ are denoted by $X$ and $Y$, respectively, in Ref.~\cite{PreviousWork}.}
	Interestingly, our current result indicates that $a^u_{\Lambda_c^+ \Xi_c^+}$ is enhanced by a factor of 2.2 once the CMM is removed from  the static bag.  The large correction can be seen by the fact that $a_{\Lambda_c^+ \Xi_c^+}^u$ has canonical dimension 3, and the only possible canonical dimension comes from the bag radius in the massless limit of $M_{u,d,s}$. \footnote{
		We do not include $M_Q$ here, as spectator quarks have little effects on the matrix elements. 
	} Therefore, it is expected that
	\begin{equation}
		a^u_{\Lambda_c^+ \Xi_c^+} \propto R^{-3}\,.
	\end{equation}
	This implies that $a_{\Lambda_c^+ \Xi_c^+}$ is sensitive to the bag radius.  It has been shown in Ref.~\cite{bagO} that the distance between the quarks will become smaller by around 20\% once the CMM is eliminated. This explains the large correction since $(4/5)^{-3} \approx 2$.
	Likewise, it is ready to evaluate the matrix elements of $O_{1,2}^c$ in a similar vein. 
	In the static bag model, it was found $a^c_{\Lambda_c^+ \Xi_c^+}=0.55\times 10^{-2}\, {\rm GeV{^3}}$ ~\cite{PreviousWork}. We see from Table \ref{table2} that this matrix element is enhanced by a factor of 1.7 after taking the CMM effect into account.

	The light diquark of the  antitriplet and  sextet baryons form spin-0 and spin-1 configurations, respectively. Since $O_1^u $  is a scalar, it is unable to bring  spin-1 to spin-0. Thus, we obtain
	\begin{equation}\label{40}
		\langle {\cal B}_{\overline{3}} |O_{1,2}^u | {\cal B}_6 \rangle = 0 \,. 
	\end{equation}
	As shown in Ref. \cite{Cheng:HFC}, the combined heavy quark and chiral symmetries severely restrict the weak transitions allowed. It turns out that ${\cal B}_{\bar 3}-{\cal B}_6$ weak transition via the weak Hamiltonian ${\cal H}^u_{\rm eff}$ is prohibited in the heavy quark limit,  namely,
	$\langle {\cal B}_{\bar 3}|{\cal H}^u_{\rm eff}|{\cal B}_6\rangle=0$, which is consistent with Eq.~\eqref{40}. 
	However, the same argument is not applicable to  $O_{1,2}^c$ since neither $cs$ nor $cd$ forms a spin eigenstate.

	The results of the matrix elements $a_{{\cal B}{\cal B}'}^{u,c}$  are summarized in  Table~\ref{table2}. For charm baryons, we take the parameters in Eq.~\eqref{input}, whereas we take $M_b=4.78$~GeV for bottom baryons. 
	
	\begin{table}[t]
		
		\caption{ Parity-conserving matrix elements of the 4-quark operators in units of $10^{-2}$ GeV$^{3}$. }\label{table2}
		\vskip 0.2cm
		\begin{tabular}{lccccccccc}
			\hline
			\hline
			$( {\cal B}, {\cal B}' ) $	& $( \Lambda_c^+, \Xi_c^+ ) $ 	& $( \Lambda^0_b, \Xi_b^0 ) $&  $( \Lambda_c ^+, \Xi^{\prime +} _c    ) $ &  $( \Sigma_c^+, \Xi^{ +} _c    ) $&  $( \Sigma_c^0, \Xi^{ 0 } _c    ) $   & $( \Omega_c, \Xi_c^0   ) $& $( \Omega_c, \Xi_c^{\prime 0}    ) $ \\
			\hline
			$a_{{\cal B}{\cal B}'}^u$ &$3.74(20)$&$3.68(20)$& $0 $& $0 $& $0 $& $0 $& $0 $\\
			$a_{ {\cal B}{\cal B}'}^c$ &$0.95(5)$ &0& $ - 1.61(9)$& $ 1.61(9) $& $ 2.28(13)$ & $ -2.26(11)$ & $ -3.97 (25) $  \\
			\hline
			\hline		
		\end{tabular}
	\end{table}

	\section{Numerical results and discussions}
	Before proceeding to the numerical results, we shall follow Ref. \cite{PreviousWork} to treat $N_c$ in 	Eq. (\ref{a1a2}) as an effective parameter and take $N_c^{\rm eff}=18.8$ and 1.88 at the $\mu=m_c$ and $\mu=\bar m_b(m_b)$ scales, respectively.
	For the heavy baryon lifetimes, we shall use \cite{pdg,LHCb:lifetime}
	\begin{eqnarray} \label{lifetimeWA}
		&& \tau(\Xi_c^+)=(453\pm 5) ~{\rm fs}, \qquad\quad~~ \tau(\Xi_c^0)=(150.5\pm 1.9) ~{\rm fs}, \quad\quad~ \tau(\Omega_c^0)=(274.5\pm 12.4) ~{\rm fs},  \nonumber \\
		&& \tau(\Xi_b^0)=(1.480\pm0.030) ~{\rm ps}, \quad \tau(\Xi_b^-)=(1.572\pm0.040) ~{\rm ps},
		\quad \tau(\Omega_b^-)=(1.64^{+0.18}_{-0.17}) ~{\rm ps}.  \nonumber\\
	\end{eqnarray}	
	
	Expressions of the $S$- and $P$-wave amplitudes for various HFC decays obtained in the previous work are collected in Eqs.~\eqref{Swave} and \eqref{Pwave}. Since the matrix elements $a^u_{\cal{B}\cal{B}'}$ evaluated in the static bag model led to ${\cal B}( \Xi_b^- \to \Lambda_b^0 \pi^-) = 6.6\times 10^{-4}$, which is too small compared to the LHCb measurement (see Eq. (\ref{LHCBb}) below),  a hybrid scheme was adopted in Ref. \cite{PreviousWork} for the matrix elements of 4-quark operators, namely, the diquark model for $a^u_{\cal{B}\cal{B}'}$ and the static bag model for $a^c_{\cal{B}\cal{B}'}$.\footnote{The diquark  model is not applicable to the matrix elements $a^c_{\cal{B}\cal{B}'}$.}
	We have noticed in passing that the matrix elements $a^u_{\cal{B}\cal{B}'}$ evaluated in the homogeneous bag model without the CMM  is consistent with the diquark model. Thus we shall use the homogeneous bag model to compute both $a^u_{\cal{B}\cal{B}'}$ and $a^c_{\cal{B}\cal{B}'}$ for reason of consistency. 
	Using the values of matrix elements  from Table~\ref{table2} and form factors from Table \ref{table1}, the calculated 
	$S$- and $P$-wave amplidutes,
	branching fractions and up-down asymmetries of various HFC decays are displayed in Table~\ref{table3}. 
	

		
	
	\begin{table}[t]
		
		\caption{ 
			The magnitudes of the $S$- and $P$-wave amplitudes (in  units of $10^{-7}$),	
			branching fractions (in units of $10^{-3}$) and the up-down asymmetries of various HFC decays. }\label{table3}
		\vskip 0.2cm
		\begin{tabular*}{0.75\textwidth}{l@{\extracolsep{\fill}} cc|c c}
			\hline
			\hline
			Mode & $A$ & $B$   & $BF$ & $\alpha$ \\
			\hline
			$\Xi_c^0 \to \Lambda_c^+\pi^-$&$-3.24\pm 0.20$ & $ -522\pm 30$~~ &$ 7.2 \pm  0.7 $  & $0.46 \pm 0.05$ \\
			$\Xi_c^+ \to \Lambda_c^+\pi^0$ &$ -2.51 \pm 0.14$&$ -410\pm 23 $~~ &$ 13.8 \pm 1.4 $  & $0.45 \pm 0.05 $\\
			$\Omega_c^0  \to \Xi_c^+\pi^-$ & ~~$ 3.06\pm 0.15 $&$ -91\pm 6 $~~ &$ 2.0 \pm 0.2 $  & $\approx -1.00$ \\
			$\Omega_c^0  \to \Xi_c^0\pi^0 $ &$ -2.17\pm 0.12 $&$ 69\pm 4 $ &$  1.1 \pm 0.1 $  & $\approx -1.00 $\\
			\hline
			$\Xi_b^- \to \Lambda_b^0\pi^-$&$ -3.27\pm 0.20 $&$ 0 $ &$4.2\pm 0. 5$ & 0 \\
			$\Xi_b^0 \to \Lambda_b^0\pi^0$&$ -2.67 \pm 0.14$&$ 0 $ &$2.6 \pm 0.3 $ & 0\\
			$\Omega_b^-  \to \Xi_b^0\pi^-$&$ 0 $&$ 16.1 \pm 0.1$ & $(6.5\pm 0.7) 10 ^{-2} $  & 0\\
			$\Omega_b^-  \to \Xi_b^-\pi^0 $ &$ 0 $&$ 3.49\pm 0.02 $ &$ (3.2 \pm 0.3)   10 ^{-3}  $ & 0 \\
			\hline
			\hline		
		\end{tabular*}
	\end{table}

	
	The first measured HFC weak decay is the $b$-flavor-conserving and strangeness-changing weak decay $\Xi_b^-\to\Lambda_b^0\pi^-$. Its relative rate was measured by LHCb to be~\cite{LHCb:HFCb}
	\beq
	{f_{\Xi_b^-}\over f_{\Lambda_b^0}} {\cal B}(\Xi_b^-\to\Lambda_b^0\pi^-)=(5.7\pm1.8^{+0.8}_{-0.9})\times 10^{-4},
	\eeq
	where $f_{\Xi_b^-}$ and $f_{\Lambda_b^0}$ are $b\to \Xi_b^-$ and $b\to\Lambda_b^0$ fragmentation fractions, respectively. The values of $f_{\Xi_b^-}/f_{\Lambda_b^0}$ were
	obtained by LHCb by invoking SU(3) symmetry in $\Xi_b^0\to J/\psi\Xi^-$ and $\Lambda_b^0\to J/\psi\Lambda$ decays  \cite{LHCb:productionratio}, leading to~\footnote{ This value can  be found explicitly in  Ref.~\cite{LHCb:HFCc}. }
	\begin{eqnarray}\label{LHCBb} 
		{\cal B}( \Xi_b^- \to \Lambda_b^0 \pi ^- )_{\text{exp}} 
		&=& (6.0 \pm 1.8)\times 10^{-3}\,.
	\end{eqnarray}
	The charm-flavor-conserving decay $\Xi_c^0\to\Lambda_c^+\pi^-$ first advocated and studied in 1992 \cite{Cheng:HFC} was finally measured by LHCb in 2021  \cite{LHCb:HFCc} and Belle very recently \cite{Belle:2022kqi} with results in excellent agreement with each other:
	\begin{eqnarray}\label{LHCBc}
		{\cal B}(\Xi_c^0 \to \Lambda_c^+ \pi ^- )_{\text{exp}} =
		\left\{ \begin{array}{c}  (0.55\pm 0.18)\% ~~{\rm LHCb},  \\  (0.54\pm0.14)\%
			~~{\rm Belle}.  \end{array} \right.
	\end{eqnarray}
	
	Our results	
	${\cal B}( \Xi_c^0 \to \Lambda_c^+ \pi^- ) = (7.2 \pm 0.7) \times 10^{-3} $ and 
	${\cal B}( \Xi_b^- \to \Lambda_b^0 \pi^- ) = (4.2 \pm 0.5) \times 10^{-3} $ are  both in agreement with experiment. The predicted branching fractions for $\Omega_b^-\to\Xi_b^0\pi^-$ and $\Omega_b^-\to\Xi_b^-\pi^0$ decays are very small as they proceed only through factorizable external and internal $W$-emission diagrams, respectively. The factorizable amplitude is proportional to the pion's momentum, which is of order 205 MeV in HFC $\Omega_b$
	decays. Consequently, their rates are suppressed. 
	
	\begin{table}
		\begin{ruledtabular}
			\caption{Branching fractions (in units of $10^{-3}$) of heavy-flavor-conserving decays $\Xi_c\to\Lambda_c\pi$ and $\Xi_b\to\Lambda_b\pi$ predicted in various models. All the model results
				have been normalized using the current world averages of lifetimes for $\Xi_c^{+,0}$ and $\Xi_b^{0,-}$ given in Eq. (\ref{lifetimeWA}). 
			} \label{table4}
			\vspace{4pt}
			\footnotesize{
				\begin{tabular}
					{ l c c c c c c c c}
					Mode  & (CLY)$^2$  & Faller & Gronau & Voloshin & Niu & HYC & This work & Exp \\
					& \cite{Cheng:HFC2016} & \cite{Faller} & \cite{Gronau:2015jgh} & \cite{Voloshin:2019} &  \cite{Niu:2021qcc} & \cite{PreviousWork} &  \\
					\hline
					$\Xi_{c}^{0}\to\Lambda_c^+\pi^-$ & $0.17$  & $<3.9$ & $0.18^{+0.23}_{-0.13}$ & $>\!0.25\pm0.15$  & $5.8\pm2.1$ & $1.76^{+0.18}_{-0.12}$ & $7.2\pm0.7$ & $5.4\pm1.1$  \\
					&   &  & $1.34\pm0.53$ \footnotemark[1]  &   &  &  &   \\
					$\Xi_{c}^{+}\to\Lambda_c^+\pi^0$ & $0.11$  & $<6.1$ & $<\!0.2$ & -- & $11.1\pm4.0$ & $3.03^{+0.29}_{-0.22}$ & $13.8\pm1.4$ & --\\
					&   & & $2.01\pm0.80$ \footnotemark[1]  & & & \\
					$\Xi_{b}^-\to\Lambda_b^0\pi^-$ & $7.0$  & $1.9-7.6$ & $6.4\pm4.3$ & $8\pm3$  & $1.4\pm0.7$ & $4.67^{+2.29}_{-1.83}$ & $4.2\pm0.5$ & $6.0\pm1.8$  \\
					$\Xi_{b}^0\to\Lambda_b^0\pi^0$ & $2.5$  & $0.9-3.7$ & $3.2\pm2.1$ & -- & $0.17\pm0.15$ & $2.87^{+1.20}_{-0.99}$ & $2.6\pm0.3$ & -- \\
				\end{tabular}
				\footnotetext[1]{Assuming (wrongly) constructive interference between the $W$-exchange diagrams induced by $cs\to dc$ and $su\to ud$ transitions.}
			}
		\end{ruledtabular}
	\end{table}

	In Table~\ref{table4}, we compare our results with other model calculations. For HFC $\Xi_c^0\to\Lambda_c^+\pi^-$ decay, it is clear that all the early predictions before 2020 with the branching fraction of order $(1\sim 3)\times 10^{-4}$ are too small compared to experiment. As first pointed out by Niu, Wang and Zhao (NWZ) \cite{Niu:2021qcc} and independenly by Groote and K\"orner  \cite{Groote:2021pxt}, the $P$-wave amplitude induced through $cs\to dc$ $W$-exchange was overlooked in all the previous model calculations. It turns out that owing to the small mass difference between $\Xi_c$ and the intermediate $\Sigma_c$ pole, of order 16 MeV, PC amplitudes are two orders of magnitude larger than the PV ones (see Table \ref{table3}). That is,  $\Xi_c\to\Lambda_c^+\pi$ receives largest contributions from the $\Sigma_c$ pole terms.
	
	We next turn to the HFC decays of the bottom baryon $\Xi_b$.  
	From Eqs. (\ref{Swave}) and (\ref{Pwave}) we see that if the factorizable contributions to
	$\Xi_b\to\Lambda_b\pi$ are neglected, we will have the relation~\cite{Gronau:2015jgh}
	\begin{equation} \label{AmpXib}
		\sqrt{2}\,A^{\Xi_b^0 \to \Lambda_b ^0\pi^0} = A^{\Xi_b^- \to \Lambda_b ^0\pi^-} \,,
	\end{equation}
	for $S$-wave amplitudes, while $P$-waves vanish, leading to 
	\begin{equation} \label{BFofXib}
		2\,{\cal B}(\Xi_b^0 \to \Lambda_b ^0\pi^0 ) \approx {\cal B}(\Xi_b^- \to \Lambda_b ^0\pi^- ) \,.
	\end{equation}
	Since the factorizable amplitude is proportional to the pion's momentum, it vanishes in the soft-pion limit. In reality, it is a small contribution to HFC decays as the pion is soft. Hence, the above relation (\ref{BFofXib}) is expected to be approximately valid. However, it is badly broken in the work of NWZ~\cite{Niu:2021qcc}, see Table \ref{table4}. To see this,  from Table VII of Ref. \cite{Niu:2021qcc} we see that  negative-parity baryon pole contributions to the $S$-wave amplitudes do respect the relation  (\ref{AmpXib}). However,  NWZ claimed a large factorizable contribution to $\Xi_b^- \to \Lambda_b ^0\pi^-$ from the external $W$-emission\footnote{On the contrary, the factorizable internal $W$-emission contribution to $\Xi_b^0\to\Lambda_b^0\pi^0$ was not considered in Ref.~\cite{Niu:2021qcc}.}
	comparable to the pole terms, which breaks the relation (\ref{AmpXib}) badly.  In our case, nonfactorizable terms dominate the $S$-wave, namely, $a^u_{\Lambda_b^0\Xi_b^0}\gg f_\pi^2(m_{\Xi_b}-m_{\Lambda_b})$, see Eq. (\ref{Swave}). At any rate, this issue will be clarified by measuring ${\cal B}(\Xi_b^0\to\Lambda_b^0\pi^0)$. 
	
	In the charm sector, $\Omega_c\to \Xi_c\pi$ acquire additional contributions from nonspectator $W$-exchange for both PC and PV amplitudes. The $P$-wave amplitude of  $\Omega_c\to \Xi_c\pi$ is enhanced by the ${\Xi'}_c$ pole, though it is not so dramatic as in the case of $\Xi_c\to\Lambda_c\pi$. The predicted branching fraction is of order $2\times 10^{-3}$ for $\Omega_c^0\to\Xi_c^+\pi^-$ and $1\times 10^{-3}$ for $\Omega_c^0\to\Xi_c^0\pi^0$. These HFC decays are accessible to LHCb, Belle and Belle II. 
	Early crude estimates given in \cite{Faller} indicated ${\cal B}(\Omega_b\to \Xi_b\pi)\sim {\cal O}(10^{-6})$ and ${\cal B}(\Omega_c\to \Xi_c\pi)< {\cal  O}(10^{-6})$.

	The asymmetry parameter $\alpha$ vanishes in the decays $\Xi_b\to \Lambda_b\pi$ and $\Omega_b\to\Xi_b\pi$ owing to the absence of $P$- and $S$-wave transitions, respectively. By contrast, it is very close to $-1$ in $\Omega_c\to\Xi_c\pi$ modes. For decays $\Xi_c^0\to\Lambda_c^+\pi^-$ and $\Xi_c^+\to\Lambda_c^+\pi^0$, the decay asymmetries are found to be positive, of order $0.45$. In the work of NWZ~\cite{Niu:2021qcc}, the corresponding decay asymmetries are $-0.16$ and $-0.007$, 
	\footnote{Decay asymmetries are obtained using the $S$- and $P$-wave amplitudes given in Table IV of Ref.~\cite{Niu:2021qcc}.} respectively. Therefore, NWZ predicted a negative up-down asymmetry in $\Xi_c^0\to\Lambda_c^+\pi^-$ and a negligible one in $\Xi_c^+\to\Lambda_c^+\pi^0$. 
	
	In short, although this work and the model of NWZ yield similar branching fractions for $\Xi_c\to \Lambda_c\pi$, we differ in the treatment of $S$-wave amplitudes: While NWZ rely on the pole model to consider negative-parity baryon pole contributions, we appeal to current algebra thanks to the soft nature of the pion produced 
	in HFC decays. Furthermore, NWZ claimed a sizable factroizable contribution to the $S$-wave
	while it is small in our case due to the soft momentum of the pion. Consequently, NWZ predict
	(i) a negative decay asymmetry in $\Xi_c^0\to\Lambda_c^+\pi^-$ and a negligible one in $\Xi_c^+\to\Lambda_c^+\pi^0$, and (ii) ${\cal B}(\Xi_b^0\to\Lambda_b^0\pi^0)\ll  {\cal B}(\Xi_b^-\to\Lambda_b^0\pi^-)$. Hence, measurements of decay asymmetries in $\Xi_c\to\Lambda_c\pi$ and the rate of $\Xi_b^0\to\Lambda_b^0\pi^0$ relative to $\Xi_b^-\to\Lambda_b^0\pi^-$ will allow to discriminate different models.

	

	\section{Conclusion}
	We have improved the numerical estimations of the previous work~\cite{PreviousWork}, where the static bag was employed. The wave functions from the homogeneous bag model are adopted in order to remove the CMM of the static bag.
	The calculations have been carried out under the same framework, and 
	we have shown that the matrix elements of 4-quark operators are enhanced about twice. 
	We have found that 
	${\cal B}( \Xi_c^0 \to \Lambda_c^+ \pi^- ) = (7.2 \pm 0.7) \times 10^{-3} $ and 
	${\cal B}( \Xi_b^- \to \Lambda_b^0 \pi^- ) = (4.2 \pm 0.5) \times 10^{-3} $,  both are in agreement with experiment.  For the yet-to-be-measured heavy-flavor-conserving decays, we find
	${\cal B}( \Xi_c ^+ \to \Lambda_c ^+\pi^0) = (13.8 \pm 1.4)\times 10^{-3}$, ${\cal B}( \Xi_b^0 \to \Lambda^0_b \pi^0) = (2.6 \pm 0.3)\times 10^{-3}$, 
	${\cal B}( \Omega_c ^0 \to \Xi_c^+\pi^-) = (2.0 \pm 0.2)\times 10^{-3}$, and ${\cal B}( \Omega_c ^0 \to \Xi_c^0\pi^0)= (1.1 \pm 0.1)\times 10^{-3}$. They are all accessible to LHCb, Belle and Belle II and can be tested in the near future.

	
	\begin{acknowledgments}
		We would like to thank Chao-Qiang Geng and Long-Ke Li for valuable discussions.
		This research was supported in part by the Ministry of Science and Technology of R.O.C. under Grant No. MOST-110-2112-M-001-025, 
		the National Natural Science Foundation of China
		under Grant Nos. U1932104, 12142502, 12147103, and the Guangdong
		Provincial Key Laboratory of Nuclear Science with No. 2019B121203010.
	\end{acknowledgments}

	
	\appendix

	\section{Baryon wave functions}
	
	For antitriplet heavy baryons, the light qaurks $(u,d,s)$ form  spin-0 configuration.  Their wave functions are given by
	\begin{eqnarray}
		&&	|\Xi_c^+,\updownarrow \rangle = \int\frac{1}{\sqrt{6} } \epsilon^{\alpha \beta \gamma} u _{a\alpha}^\dagger(\vec{x}_1) s_{b\beta}^\dagger(\vec{x}_2) c_{c\gamma}^\dagger (\vec{x}_3) \Psi^{abc}_{A_\updownarrow(usc)} (\vec{x}_1,\vec{x}_2,\vec{x}_3) [d^3  \vec{x}] | 0\rangle\nonumber\,,\\
		&&	|\Xi_c^0 ,\updownarrow\rangle = \int\frac{1}{\sqrt{6} } \epsilon^{\alpha \beta \gamma} d _{a\alpha}^\dagger(\vec{x}_1) s_{b\beta}^\dagger(\vec{x}_2) c_{c\gamma}^\dagger (\vec{x}_3) \Psi^{abc}_{A_\updownarrow(dsc)} (\vec{x}_1,\vec{x}_2,\vec{x}_3) [d^3  \vec{x}] | 0\rangle\,,\\
		&&	|\Lambda_c^+, \updownarrow\rangle = \int\frac{1}{\sqrt{6} } \epsilon^{\alpha \beta \gamma} d _{a\alpha}^\dagger(\vec{x}_1) u_{b\beta}^\dagger(\vec{x}_2) c_{c\gamma}^\dagger (\vec{x}_3) \Psi_{A_\updownarrow(duc)}^{abc} (\vec{x}_1,\vec{x}_2,\vec{x}_3) [d^3  \vec{x}] | 0\rangle\,,\nonumber
	\end{eqnarray}
	where the spatial parts of the wave functions are described by Eq.~\eqref{xdelta}.

	On the other hand, 
	the light quarks of 
	the sextet baryons form spin-1 configuration. Their wave functions read
	\begin{eqnarray}
		&&	|\Sigma_c^{  ++}  , \updownarrow\rangle = \int\frac{1}{2\sqrt{3} } \epsilon^{\alpha \beta \gamma} u _{a\alpha}^\dagger(\vec{x}_1) u_{b\beta}^\dagger(\vec{x}_2) c_{c\gamma}^\dagger (\vec{x}_3) \Psi^{abc}_{S_\updownarrow (uuc)} (\vec{x}_1,\vec{x}_2,\vec{x}_3) [d^3  \vec{x}] | 0\rangle\,,\nonumber\\
		&&	|\Sigma_c^{  +}  , \updownarrow \rangle = \int\frac{1}{\sqrt{6} } \epsilon^{\alpha \beta \gamma} d _{a\alpha}^\dagger(\vec{x}_1) u_{b\beta}^\dagger(\vec{x}_2) c_{c\gamma}^\dagger (\vec{x}_3) \Psi^{abc}_{S_\updownarrow (duc)} (\vec{x}_1,\vec{x}_2,\vec{x}_3) [d^3  \vec{x}] | 0\rangle\,,\nonumber\\
		&&	|\Sigma_c^{  0 }  , \updownarrow \rangle = \int\frac{1}{2 \sqrt{3} } \epsilon^{\alpha \beta \gamma} d _{a\alpha}^\dagger(\vec{x}_1) d_{b\beta}^\dagger(\vec{x}_2) c_{c\gamma}^\dagger (\vec{x}_3) \Psi^{abc}_{S_\updownarrow (ddc)} (\vec{x}_1,\vec{x}_2,\vec{x}_3) [d^3  \vec{x}] | 0\rangle\,,\nonumber\\
		&&	|\Xi_c^{ \prime +}  , \updownarrow \rangle = \int\frac{1}{\sqrt{6} } \epsilon^{\alpha \beta \gamma} u _{a\alpha}^\dagger(\vec{x}_1) s_{b\beta}^\dagger(\vec{x}_2) c_{c\gamma}^\dagger (\vec{x}_3) \Psi^{abc}_{S_\updownarrow (usc)} (\vec{x}_1,\vec{x}_2,\vec{x}_3) [d^3  \vec{x}] | 0\rangle\,,\nonumber\\
		&&	|\Xi_c^{\prime 0} , \updownarrow \rangle = \int\frac{1}{\sqrt{6} } \epsilon^{\alpha \beta \gamma} d _{a\alpha}^\dagger(\vec{x}_1) s_{b\beta}^\dagger(\vec{x}_2) c_{c\gamma}^\dagger (\vec{x}_3) \Psi^{abc}_{S_\updownarrow (dsc)} (\vec{x}_1,\vec{x}_2,\vec{x}_3) [d^3  \vec{x}] | 0\rangle\,,\nonumber\\
		&&	|\Omega_c^{ 0}, \updownarrow \rangle = \int\frac{1}{2 \sqrt{3} } \epsilon^{\alpha \beta \gamma} s _{a\alpha}^\dagger(\vec{x}_1) s_{b\beta}^\dagger(\vec{x}_2) c_{c\gamma}^\dagger (\vec{x}_3) \Psi^{abc}_{S_\updownarrow (ssc)} (\vec{x}_1,\vec{x}_2,\vec{x}_3) [d^3  \vec{x}] | 0\rangle\,,
	\end{eqnarray}
	where 
	\begin{equation}
		\begin{aligned}
			\Psi^{abc} _{S_\uparrow} ( \vec{x}_1,\vec{x}_2,\vec{x}_3 ) &= \frac{{\cal N}}{\sqrt{6}} \int \left( 2\phi^a_{q_1\uparrow}(\vec{x}_1') \phi^b_{q_2\uparrow}(\vec{x}_2') \phi^c_{q_3\downarrow}(\vec{x}_3') \right. \\
			&\left. - \phi^a_{q_1\uparrow}(\vec{x}_1' ) \phi^b_{q_2\downarrow}(\vec{x}_2') \phi^c_{q_3\uparrow}(\vec{x}_3') 
			- \phi^a_{q_1\downarrow}(\vec{x}_1') \phi^b_{q_2\uparrow}(\vec{x}_2') \phi^c_{q_3\uparrow}(\vec{x}_3') \right)d^3 \vec{x}_\Delta \,,
		\end{aligned}\nonumber 
	\end{equation}
	\begin{equation}
		\begin{aligned}
			\Psi^{abc} _{S_\downarrow}  (\vec{x}_1,\vec{x}_2,\vec{x}_3 ) &=  \frac{{\cal N}}{\sqrt{6}} \int \left(  -  2\phi^a_{q_1\downarrow}(\vec{x}'_1) \phi^b_{q_2\downarrow}(\vec{x}'_2) \phi^c_{q_3\uparrow}(\vec{x}'_3) \right. \\
			&\left. + \phi^a_{q_1\downarrow}(\vec{x}'_1 ) \phi^b_{q_2\uparrow}(\vec{x}'_2) \phi^c_{q_3\downarrow}(\vec{x}'_3) 
			+  \phi^a_{q_1\uparrow}(\vec{x}'_1) \phi^b_{q_2\downarrow}(\vec{x}'_2) \phi^c_{q_3\downarrow}(\vec{x}'_3) \right)d^3 \vec{x}_\Delta \,,
		\end{aligned}
	\end{equation}
	with $\vec{x}'_i = \vec{x}_i  - \vec{x}_\Delta$. 
	
	The wave functions of the $b$-baryons can be  obtained by substituting $b$ for $c$.

	\section{Evaluation of $a^u_{\Lambda_c^+ \Xi_c^+}$}	
	In this appendix we evaluate one of the matrix elements, say $a^u_{\Lambda_c^+ \Xi_c^+}$, explicitly.
	We start with sorting out the overlapping of the spectator quark
	\begin{eqnarray}\label{col0}
		{\cal D}_c(\vec{x}_\Delta)  &=& \int d^3\vec{x}
		\left(  \begin{array}{cc}
			u_c^+ \chi^\dagger , & -i v_c^+ \chi^\dagger\sigma_{x^+} 
		\end{array}\right )\cdot 
		\left(  \begin{array}{c}
			u_c^- \chi \\
			i v_c^- \sigma_{x^-} \chi 
		\end{array}\right )\,,\nonumber\\
		&=&\int d^3\vec{x}\left[  u_c^+ u_c^- + v_c^+v_c^- \chi^\dagger \left( 
		\sigma_{x^+}  \sigma_{x^-} 
		\right) \chi \right] \,,
		\nonumber\\
		&=&  \int d^3\vec{x}\left[
		u_c^+ u_c^- + v_c^+v_c^-  \left( 
		\hat{x}^+ \cdot \hat{x}^- +\frac{1}{r^+r^- } i  \chi^\dagger  (\vec{x}_\Delta \times \vec{x}) \cdot \vec{\sigma}\chi
		\right)  \right]\,,
		\nonumber\\
		&=&  \int d^3\vec{x}\left[
		u_c^+ u_c^- + v_c^+v_c^-  \left( 
		\hat{x}^+ \cdot \hat{x}^- 
		\right)  \right] \,,
	\end{eqnarray}	
	where   we have taken the shorthand 
	$
	\sigma _{{v}} = \hat{v} \cdot \vec{\sigma}$ with $\hat{v}$ an arbitrary unit vector, $u_q^\pm = u_q(r^\pm)$ and $v_q^\pm = v_q(r^\pm)$ with $r^\pm = | \vec{x}^\pm|$.  To get the third line of the equation, we have used the relation $\sigma_i \sigma_j = \delta_{ij} + i \epsilon^{ijk}\sigma_k$, 

	The integrals in $\Gamma(\vec{x}_\Delta)$ are much more complicated. 
	We decompose them into several parts as  
	\begin{equation}\label{col1}
		\Gamma(\vec{x}_\Delta)  = \int d^3\vec{x} \sum_{i=1,2,3,4} \Gamma_i(\vec{x}_\Delta,\vec{x}) +\text{PV} 
	\end{equation}
	with 
	\begin{eqnarray}\label{B3}
		\Gamma_1(\vec{x}_\Delta,\vec{x})  &=&
		\sum_{[\lambda]}  \frac{1}{2}	\left(
		-1
		\right)^{\lambda_1- \lambda_3} 
		\phi _{d\lambda_4} ^\dagger\left(\vec{x}^ + \right) \phi_{u\lambda_2}\left(\vec{x}^- \right) \phi _{u\lambda_3} ^\dagger\left(\vec{x} ^+ \right) \phi_{s\lambda_1}\left(\vec{x} ^- \right) \,, \nonumber\\
		\Gamma_2(\vec{x}_\Delta,\vec{x})&=&
		\sum_{[\lambda]}  \frac{1}{2}	\left(
		-1
		\right)^{\lambda_1- \lambda_3} 
		\phi _{d\lambda_4} ^\dagger\left(\vec{x}^ + \right) \gamma_5 \phi_{u\lambda_2}\left(\vec{x}^- \right) \phi _{u\lambda_3} ^\dagger\left(\vec{x} ^+ \right) \gamma_5 \phi_{s\lambda_1}\left(\vec{x} ^- \right)  \,,\nonumber\\
		\Gamma_3(\vec{x}_\Delta,\vec{x})&=&-
		\sum_{[\lambda]}  \frac{1}{2}	\left(
		-1
		\right)^{\lambda_1- \lambda_3} 
		\phi _{d\lambda_4} ^\dagger\left(\vec{x}^ + \right) V_i \phi_{u\lambda_2}\left(\vec{x}^- \right) \phi _{u\lambda_3} ^\dagger\left(\vec{x} ^+ \right) V_i \phi_{s\lambda_1}\left(\vec{x} ^- \right) \,, \\
		\Gamma_4(\vec{x}_\Delta,\vec{x})&=& -
		\sum_{[\lambda]}  \frac{1}{2}	\left(
		-1
		\right)^{\lambda_1- \lambda_3} 
		\phi _{d\lambda_4} ^\dagger\left(\vec{x}^ + \right) V_i\gamma_5 \phi_{u\lambda_2}\left(\vec{x}^- \right) \phi _{u\lambda_3} ^\dagger\left(\vec{x} ^+ \right) V_i \gamma_5\phi_{s\lambda_1}\left(\vec{x} ^- \right)  , \nonumber
	\end{eqnarray}
	where $V _i = \gamma_0 \gamma_i$ with $i=1,2,3$ and PV stands for the parity-violating part, which is not concerned in this work.

	In tidying up the spin indices, it is useful to note the identities
	\begin{equation}\label{B4}
		\begin{aligned}
			&		\sum_{[\lambda]}\frac{1}{2} \left(
			-1
			\right)^{\lambda_1- \lambda_3}  ( \chi_{\lambda_3}^\dagger  \chi_{\lambda_1} )( \chi_{\lambda_4} ^\dagger \chi_{\lambda_2}) =  1  \,,\\
			&		\sum_{[\lambda]}\frac{1}{2} \left(
			-1
			\right)^{\lambda_1- \lambda_3}  ( \chi_{\lambda_3}^\dagger  \sigma_i \chi_{\lambda_1} )( \chi_{\lambda_4} ^\dagger  \chi_{\lambda_2}) = \sum_{[\lambda]}\frac{1}{2} \left(
			-1
			\right)^{\lambda_1- \lambda_3}  ( \chi_{\lambda_3}^\dagger  \chi_{\lambda_1} )( \chi_{\lambda_4} ^\dagger\sigma_i  \chi_{\lambda_2}) =  0   \,,\\
			& \sum_{[\lambda]}\frac{1}{2} \left(
			-1
			\right)^{\lambda_1- \lambda_3}  ( \chi^\dagger_{\lambda_3} \sigma_i \chi_{\lambda_1} )( \chi_{\lambda_4}^\dagger \sigma_j \chi_{\lambda_2}) =  -\delta_{ij} \,,
		\end{aligned} 
	\end{equation}
	which can be derived 
	from a direct calculation. It can be interpreted as the interacting quarks are spin-0,  so the matrix elements cannot depend on a specific direction. 
	Using Eq.~\eqref{B4}, we arrive at 
	\begin{eqnarray}\label{col2}
		\Gamma_1(\vec{x}_\Delta,\vec{x}) &=&
		\left( 
		u_d^+ u_u^- +  v_d^+v_u^-  
		\hat{x}^+ \cdot \hat{x}^-  
		\right)\left( 
		u_u^+ u_s^- +  v_u^+v_s^-  
		\hat{x}^+ \cdot \hat{x}^-   
		\right) + \frac{(\vec{x}_\Delta \times \vec{x}  ) ^2}{(r^+r^-)^2}  v_d^+v_u^-   v_u^+v_s^- \,, \nonumber\\
		\Gamma_2(\vec{x}_\Delta , \vec{x})  &=& \left(   u_d^+v_u^- \hat{x}^- -  v_d^+  u_u^-  \hat{x}^+ \right) 
		\left( u_u^+  v_s^-  \hat{x}^- -  v_u^+  u_s^-  \hat{x}^+ \right) \,,\nonumber\\
		\Gamma_3(\vec{x}_\Delta , \vec{x})  &=&  \Gamma_2(\vec{x}_\Delta, \vec{x}) 
		+ 2\left( u_d^+  v_u^-  \hat{x}^- +  v_d^+  u_u^-  \hat{x}^+ \right) \cdot
		\left( u_u^+  v_s^-  \hat{x}^- +  v_u^+  u_s^-  \hat{x}^+ \right)  \,,\nonumber\\
		\Gamma_4(\vec{x}_\Delta , \vec{x})  &=&   3 u_d^+u_u^- u_u^+ u_s^- +
		v_d^+v_u^- v_u^+ v_s^-  \left( 2 +(\hat{x}^+ \cdot\hat{x}^-)^2 +\frac{   (\vec{x}_\Delta \times \vec{x}) ^2 }{(r^+r^-)^2}
		\right) \nonumber\\
		&& -(u_d^+u_u^- v_u^+ v_s^- + v_d^+v_u^- u_u^+ u_s^-  ) \hat{x}^+\cdot \hat{x}^-\,.
	\end{eqnarray}
	To see the connection with the static limit, we take $\vec{x}_\Delta = 0 $ and obtain
	\begin{equation}
		a^u_{\Lambda_c^+ \Xi_c^+} = 2 \int \sum_{k = 1,2,3,4} \Gamma_k ( 0,\vec{x})d^3\vec{x} = 2 \int 4(u_du_u + v_dv_u) ( u_u u_s +v_uv_s) ( 4\pi r^2 dr)\,.
	\end{equation}
	This is precisely the quantity $X$ defined in Eq.~(4.1) of Ref.~\cite{PreviousWork}.

	The integrals of $d^3\vec{x}$ in Eqs.~\eqref{col0} and \eqref{col1} can be further simplified by adopting the cylindrical coordinate with $\vec{x}_\Delta$ in the $z$ direction given as 
	\begin{eqnarray}
		&&{\cal D}(\vec{x}_\Delta) 
		= 2\pi \int^{\sqrt{R^2 - r_\Delta^2/4 }}_0 \rho d\rho 
		\int^{\sqrt{R^2 - \rho^2 } - r_\Delta /2 }_{-\sqrt{R^2 - \rho^2 } + r_\Delta /2}
		d z 
		\left(
		u_c^+ u_c^- + v_c^+ v_c^- \frac{4\rho^2 +4 z^2 - r_\Delta ^2 }{4 r^+r^- }
		\right)\,,
		\nonumber\\
		&&\Gamma(\vec{x}_\Delta) =  2\pi 
		\int^{\sqrt{R^2 - r_\Delta^2/4 }}_0  \rho d\rho 
		\int^{\sqrt{R^2 - \rho^2 } - r_\Delta /2 }_{-\sqrt{R^2 - \rho^2 } + r_\Delta /2}
		d z  \sum \Gamma_i(\vec{x}_\Delta, \rho  , z)\,,
	\end{eqnarray}
	where
	$r_\Delta = | \vec{x}_\Delta| $\,, and
	we have used that the integrands are independent of the azimuthal angule, and the  bounds of the integrals come from the bag boundary. It is easily seen that ${\cal D}(\vec{x}_\Delta)$ and $\Gamma(\vec{x}_\Delta)$
	depend only on the magnitude of $\vec{x}_\Delta$. 
	Thus, Eq.~\eqref{O4Ex} can be recast as 
	\begin{equation}
		\int d^3 \vec{x}_\Delta  {\cal D}_c(\vec{x}_\Delta) \Gamma (\vec{x}_\Delta) = 4 \pi \int^{2R}_0 dr_\Delta  {\cal D}_c(r_\Delta) \Gamma (r_\Delta)\,.
	\end{equation}
	Here the upper limit of $r_\Delta$ comes from that the static bags separated farther than $2R$ do not overlap.
	This completes the evaluation of $a^u_{\Lambda_c^+ \Xi_c^+}$.


\vskip 1.5cm
	\begin{thebibliography}{99}
		
		\bibitem{Cheng:HFC}
		H.~Y.~Cheng, C.~Y.~Cheung, G.~L.~Lin, Y.~C.~Lin, T.~M.~Yan and H.~L.~Yu,
		``Heavy flavor conserving nonleptonic weak decays of heavy baryons'',
		Phys.\ Rev.\ D {\bf 46}, 5060 (1992).
		
		
		\bibitem{PreviousWork}
		H.~Y.~Cheng and F.~Xu,
		``Heavy-flavor-conserving hadronic weak decays of charmed and bottom baryons,''
		Phys. Rev. D \textbf{105}, no.9,  094011 (2022)
		[arXiv:2204.03149 [hep-ph]].
		
		
		\bibitem{LHCb:HFCc}
		R.~Aaij \textit{et al.} [LHCb],
		``First branching fraction measurement of the suppressed decay $\Xi_c^0\to \pi^-\Lambda_c^+$,''
		Phys. Rev. D \textbf{102}, no.7, 071101 (2020).
		[arXiv:2007.12096 [hep-ex]].
		
		\bibitem{Liu:2022pdk}
		C.~W.~Liu and C.~Q.~Geng,
		``Center of mass motion in bag model,''
		[arXiv:2205.08158 [hep-ph]].
		
		\bibitem{pdg}
		R.~L.~Workman \textit{et al.} [Particle Data Group],
		``Review of Particle Physics,''
		PTEP \textbf{2022}, 083C01 (2022).
		
		
		\bibitem{Zou:2019kzq}
		J.~Zou, F.~Xu, G.~Meng and H.~Y.~Cheng,
		``Two-body hadronic weak decays of antitriplet charmed baryons,''
		Phys. Rev. D \textbf{101}, no.1, 014011 (2020)
		[arXiv:1910.13626 [hep-ph]].
		
		
		\bibitem{Singly Cabibbo}
		H.~Y.~Cheng, X.~W.~Kang and F.~Xu,
		``Singly Cabibbo-suppressed hadronic decays of $\Lambda_c^+$,''
		Phys. Rev. D \textbf{97}, no.7, 074028 (2018)
		[arXiv:1801.08625 [hep-ph]].
		
		\bibitem{bagO}
		C.~Q.~Geng, C.~W.~Liu and T.~H.~Tsai,
		``Nonleptonic two-body weak decays of $\Lambda_b$ in modified MIT bag model,''
		Phys. Rev. D \textbf{102}, no.3, 034033 (2020)
		[arXiv:2007.09897 [hep-ph]].
		
		
		\bibitem{DeGrand:1975cf}
		A.~Chodos, R.~L.~Jaffe, K.~Johnson and C.~B.~Thorn,
		``Baryon Structure in the Bag Theory,''
		Phys. Rev. D \textbf{10}, 2599 (1974);
		T.~A.~DeGrand, R.~L.~Jaffe, K.~Johnson and J.~E.~Kiskis,
		``Masses and Other Parameters of the Light Hadrons,''
		Phys. Rev. D \textbf{12}, 2060 (1975).
		
		\bibitem{Korner:1970xq}
		J.~G.~Korner,
		``Octet behaviour of single-particle matrix elements $ < B'|H(W)|B > $ and $< M'|H(W)|M>$ using a weak current current quark Hamiltonian,''
		Nucl. Phys. B \textbf{25}, 282 (1971);
		J.~C.~Pati and C.~H.~Woo,
		``$\Delta I$ = 1/2 rule with fermion quarks,''
		Phys. Rev. D \textbf{3}, 2920 (1971).
		
		
		
		\bibitem{Diquark}
		H.~Y.~Cheng, C.~Y.~Cheung, G.~L.~Lin, Y.~C.~Lin, T.~M.~Yan and H.~L.~Yu,
		``Heavy-Flavor-Conserving Hadronic Weak Decays of Heavy Baryons,''
		JHEP \textbf{03}, 028 (2016)
		[arXiv:1512.01276 [hep-ph]].
		
		
		
		\bibitem{LHCb:lifetime}
		R.~Aaij \textit{et al.} [LHCb Collaboration],
		``Measurement of the lifetimes of promptly produced $\Omega^{0}_{c}$ and $\Xi^{0}_{c}$ baryons,''
		Sci. Bull. \textbf{67}, 479-487 (2022)
		[arXiv:2109.01334 [hep-ex]].
		
		\bibitem{Cheng:HFC2016}
		H.~Y.~Cheng, C.~Y.~Cheung, G.~L.~Lin, Y.~C.~Lin, T.~M.~Yan and H.~L.~Yu,
		``Heavy-Flavor-Conserving Hadronic Weak Decays of Heavy Baryons,''
		JHEP \textbf{03}, 028 (2016)
		[arXiv:1512.01276 [hep-ph]].
		
		\bibitem{Faller}
		S.~Faller and T.~Mannel,
		``Light-Quark Decays in Heavy Hadrons,''
		Phys. Lett. B \textbf{750}, 653-659 (2015)
		[arXiv:1503.06088 [hep-ph]].
		
		\bibitem{Gronau:2015jgh}
		M.~Gronau and J.~L.~Rosner,
		``$S$-wave nonleptonic hyperon decays and $\Xi^-_b \to \pi^- \Lambda_b$,''
		Phys. Rev. D \textbf{93}, no.3, 034020 (2016)
		[arXiv:1512.06700 [hep-ph]];
		M.~Gronau and J.~L.~Rosner,
		``From $\Xi_b \to \Lambda_b \pi$ to $\Xi_c \to \Lambda_c \pi$,''
		Phys. Lett. B \textbf{757}, 330-333 (2016)
		[arXiv:1603.07309 [hep-ph]].
		
		\bibitem{Voloshin:2019}
		M.~B.~Voloshin,
		``Update on splitting of lifetimes of $c$ and $b$ hyperons within the heavy quark expansion and decays $\Xi_Q \to \Lambda_Q \pi$,''
		Phys. Rev. D \textbf{100}, no.11, 114030 (2019)
		[arXiv:1911.05730 [hep-ph]].
		
		\bibitem{Niu:2021qcc}
		P.~Y.~Niu, Q.~Wang and Q.~Zhao,
		``Study of heavy quark conserving weak decays in the quark model,''
		Phys. Lett. B \textbf{826}, 136916 (2022)
		[arXiv:2111.14111 [hep-ph]].
		
		
		
		
		
		\bibitem{LHCb:HFCb}
		R.~Aaij \textit{et al.} [LHCb],
		``Evidence for the strangeness-changing weak decay $\Xi_b^-\to\Lambda_b^0\pi^-$,''
		Phys. Rev. Lett. \textbf{115}, no.24, 241801 (2015)
		[arXiv:1510.03829 [hep-ex]].
		
		
		\bibitem{LHCb:productionratio}
		R.~Aaij \textit{et al.} [LHCb],
		``Measurement of the mass and production rate of $\Xi_b^-$ baryons,''
		Phys. Rev. D \textbf{99}, no.5, 052006 (2019)
		[arXiv:1901.07075 [hep-ex]].
		
		
		\bibitem{Belle:2022kqi}
		[Belle],
		``Measurement of the branching fraction of $\Xi_{c}^{0}\to \Lambda_{c}^{+}\pi^{-}$ at Belle,''
		[arXiv:2206.08527 [hep-ex]].
		
		
		
		
		
		\bibitem{Groote:2021pxt}
		S.~Groote and J.~G.~K\"orner,
		``Topological tensor invariants and the current algebra approach: Analysis of 196 nonleptonic two-body decays of single and double charm baryons - a review,''
		Eur. Phys. J. C \textbf{82} (2022) 297
		[arXiv:2112.14599 [hep-ph]].
		
		
		
	\end{thebibliography}
\end{document}